\documentclass[sigconf]{acmart}

\usepackage{balance}

\def\BibTeX{{\rm B\kern-.05em{\sc i\kern-.025em b}\kern-.08emT\kern-.1667em\lower.7ex\hbox{E}\kern-.125emX}}

\usepackage{booktabs} 
\usepackage{algorithm,algpseudocode}
\usepackage{graphicx} 
\usepackage{subcaption}
\usepackage{mathtools}
\usepackage{multirow}
\usepackage[inline]{enumitem}
\usepackage{xcolor}

\setcopyright{rightsretained}

\theoremstyle{remark}

\theoremstyle{theorem}

\newcommand{\Ib}{\mathbf{I}}

\copyrightyear{2019} 
\acmYear{2019} 
\setcopyright{acmlicensed}
\acmConference[KDD '19]{The 25th ACM SIGKDD Conference on Knowledge Discovery and Data Mining}{August 4--8, 2019}{Anchorage, AK, USA}
\acmBooktitle{The 25th ACM SIGKDD Conference on Knowledge Discovery and Data Mining (KDD '19), August 4--8, 2019, Anchorage, AK, USA}
\acmPrice{15.00}
\acmDOI{10.1145/3292500.3330764}
\acmISBN{978-1-4503-6201-6/19/08}

\begin{document}
\title{Feedback Shaping: A Modeling Approach to \\Nurture Content Creation}

\author{Ye Tu}
\affiliation{\institution{LinkedIn Corporation}}
\email{ytu@linkedin.com}

\author{Chun Lo}
\affiliation{\institution{LinkedIn Corporation}}
\email{chunlo@linkedin.com}

\author{Yiping Yuan}
\affiliation{\institution{LinkedIn Corporation}}
\email{ypyuan@linkedin.com}

\author{Shaunak Chatterjee}
\affiliation{\institution{LinkedIn Corporation}}
\email{shchatterjee@linkedin.com}

\renewcommand{\shortauthors}{Tu et al.}

\begin{abstract}
Social media platforms bring together content creators and content consumers through recommender systems like newsfeed. The focus of such recommender systems has thus far been primarily on modeling the content consumer preferences and optimizing for their experience. However, it is equally critical to nurture content creation by prioritizing the creators' interests, as quality content forms the seed for sustainable engagement and conversations, bringing in new consumers while retaining existing ones.  In this work, we propose a modeling approach to predict how feedback from content consumers incentivizes creators. We then leverage this model to optimize the newsfeed experience for content creators by reshaping the feedback distribution, leading to a more active content ecosystem. Practically, we discuss how we balance the user experience for both consumers and creators, and how we carry out online A/B tests with strong network effects. We present a deployed use case on the LinkedIn newsfeed, where we used this approach to improve content creation significantly without compromising the consumers' experience.

\end{abstract}

%
%


\maketitle

\section{Introduction}
\label{sec:introduction}


Content ecosystems are an integral component of several social media platforms. This is true of Facebook, Instagram, LinkedIn, Twitter, and SnapChat, to name a few. In such an ecosystem, members play two roles --- one as content creators and the other as content consumers. A creator creates and publishes, or reshares various forms of content (e.g., articles, images, videos, etc.) to express her views and opinions, which add to the content liquidity for consumers who can see that creator's content. Consumers, in return, provide attention and feedback to the creators. A content ecosystem is very likely to grow when both creators and consumers are thriving.


There has been plenty of research \cite{agarwal2011click, Agarwal:2014:ARL:2623330.2623362, Agarwal:2015:PLF:2783258.2788614, gubin2017news} in recommender systems on predicting what consumers may find interesting. There are a few reasons for this focus. First, the modeling problem of identifying content items (from a candidate pool of potentially millions) is challenging, yet well-specified since the consumer behavior is observed and attributable to the recommendation. Second, the impact is fairly large if the candidate pool is big. Finally, measuring progress of modeling improvements is straightforward both via offline validation and online experiments randomized on the consumer.

Modeling approaches to improve the creator experience have been less common. To formulate the problem, it is important to enumerate certain salient points about the creator experience on such platforms. Upon sharing a piece of content, the creator's network (i.e., audience) can view and interact with the content by liking, commenting and re-sharing it. We refer to these signals from the consumers to the creator collectively as \textit{feedback}. One key motivation for content creators is to hear from their desired audience via feedback. This has been proved by past studies which demonstrate the effect of feedback as a strong social influence in online communities \cite{Hoffmann2009AmplifyingCC, eckles2016estimating} and we have also observed a positive correlation between feedback and future content creation in the LinkedIn content ecosystem (analysis in Section \ref{sec:utilityEstimations}). 

The creator is notified about the feedback via notifications and dashboards among other means, with different platforms employing slightly different means. Creator-facing dashboards facilitate insights of how their audience are engaging with their content. Occasionally, some platforms boost creators when they satisfy some special criteria. An example is a platform selectively notifying a creator's audience when the creator creates content after a long period of time. \textit{However, there is no systematic, data-driven, model-based solution to this problem to the best of our knowledge}. To build such a solution, we need to address two key challenges:
\begin{itemize}
\item Attribution of the reward (i.e., future creation) to the intervention (i.e., feedback) is difficult, as the ground truth effects of feedback to creation is unobservable.
\item Standard randomized experiments assume Stable Unit Treatment Value Assumption (SUTVA) \cite{rubin1974estimating}. However, measuring impact on creators would violate this assumption and requires a special design of the experiment. 
\end{itemize}

There is also the additional problem of unobservable confounders which influence (may be causally) creation behavior. We do not address this further in our work, but call it out as a potential risk.

Our proposed solution involves the definition of a creator side utility which quantifies the incremental creation behavior driven by an additional unit of feedback, henceforth called ``feedback sensitivity''. We propose a model-based method to estimate each creator's feedback sensitivity and then use that estimated sensitivity to modify consumer side recommendations by adding a creator side utility to the ranking or scoring function. We use this solution on the LinkedIn newsfeed and show how it improves upon the previous heuristic based approach \cite{bonnie2018}.

The key contributions of our work are as follows:
\begin{itemize}
\item Defining a new creator side utility based on feedback sensitivity and proposing a model-based approach to estimate that utility. Our proposal includes strategies to deal with the unobservable ground truth scenario in both offline validation and online experimentation scenarios.
\item Proposing and implementing a framework that uses this creator side utility in an online ranking system to increase future creation activity, by reshaping the distribution of feedback.
\item Demonstrating success from online experiments on a social media platform (i.e., LinkedIn) which benefited hundreds of thousands of creators, without compromising the consumer experience.
\end{itemize}

The paper is organized as follows: Section \ref{sec:relatedWork} discusses the related work in developing feed recommendation system; estimating unobservable utility; optimizing creator side utility; and measuring peer impacts through the network A/B experiment design. We formally describe our problem statement in Section \ref{sec:problem} and discuss how we develop our models and estimate the utilities in Section \ref{sec:utilityEstimations}. Next, we show the application of utilities in Feed recommendation system in Section \ref{sec:feedApplication} and illustrate our system architecture in section \ref{sec:architecture}. Finally, we summarize our findings and learnings along with some future work in Section \ref{sec:discussion}.

\section{Related Work}
\label{sec:relatedWork}

There have been several approaches proposed to predict model-item click behavior including item-based collaborative filtering \cite{sarwar2001item}, matrix factorization \cite{koren2009matrix} (popularized immensely by the Netflix challenge) and wide and deep recommender systems \cite{cheng2016wide}. This prediction task is a basic building block for consumer utility optimization in social media platform products like newsfeed. There are several papers describing the nuances of ranking a newsfeed --- both at LinkedIn \cite{Agarwal:2014:ARL:2623330.2623362, Agarwal:2015:PLF:2783258.2788614} and beyond \cite{li2010personalizefeed, gubin2017news}. However, all of these works focus on the consumer experience optimization.

One other related line of work is multi-objective optimization. As the products evolve, multiple objectives (e.g., member engagement, ads revenue) are driven by a complex product like newsfeed. In order to achieve the optimal trade-off among potentially conflicting objectives, constrained optimization based approaches relying on Lagrangian dual estimation have been proposed \cite{agarwal2011click} and used in multiple applications \cite{agarwal2015constrained, gupta2016email, gao2018near}. We use the same approach to introduce a creator side utility into the existing multi-objective feed ranking function.

There has been some work on improving the creator experience and incenting them to create more content. Examples include motivating newer users on Facebook to become content creators \cite{burke2009feed}, studying the impact of social ties on content creation, and vice versa \cite{shriver2013social}, increasing content creation in a non-social network setting by finding synergies between content creation and information extraction tasks \cite{Hoffmann2009AmplifyingCC}. More recently, there was a heuristic-based creator utility proposed and tested at LinkedIn \cite{bonnie2018}.

Unlike the observed response prediction problem, which is often formulated as a classification problem, the exact feedback sensitivity is not observable. A creator receives feedback from many members in her potential audience before creating again. We do not have observation data on how the creator would behave if a particular member's feedback was the last feedback (in case it isn't). One recent work studying such personalized (or fine-grained) effect estimation looks at a related problem of how individual notifications affect a member's propensity to visit a site \cite{yuan2018}. An alternate approach aimed at estimating more coarse-grained effect is observational causal study. Such studies estimate the average treatment effect of the intervention on the potential outcome \cite{aronow2017estimating, toulis2013estimation, frangakis2002principal, forastiere2016interference}. Our work follows the methodology introduced in \cite{yuan2018} to a large extent.

Finally, in order to test the benefit of introducing our creator utility in the LinkedIn feed ranking function, we need to design a special randomized test. Conventional randomized tests \cite{rubin1980randomization} (or A/B tests or bucket tests) suffice for consumer side optimization, but are inadequate to measure the impact of creators. This is due to the potential network effect of any treatment that redistributes feedback, since such a treatment violates the ``Stable Unit Treatment Value Assumption'' or SUTVA principle \cite{rubin1974estimating}. Thankfully, this has been a hot research area in the last few years \cite{katzir2012framework, toulis2013estimation, ugander2013graph, aronow2012estimating, gui2015network, kohavi2013online, eckles2016estimating, taylor2017randomized}. At LinkedIn, we currently use a modified version ~\cite{Guillaume2019} of the method described in \cite{gui2015network}, which was also used in \cite{bonnie2018}.

\section{Problem Statement}
\label{sec:problem}

A content creator on a social network is typically motivated to engage with her intended audience. Established content creators generally prefer their audience to keep growing in size, while new ones would love to get some feedback to confirm that their voice is being heard. \textit{If creators find value in the platform, they continue to create more} -- this assumption forms the basis of our choice to use creation propensity as a proxy for creator value. A model which better predicts how feedback affects a creator's future creation behavior can be effectively used as a proxy to represent creator interests during item ranking for consumers.

Our focus is to understand how feedback as an intervention can facilitate creators to create more. The estimated impact of feedback on a creator can be used as a utility to balance the interests between consumers and creators using a linear combination. It can be shown that this is also the solution to a constrained optimization problem of maximizing consumer utility while ensuring a certain amount of creator utility \cite{agarwal2012personalized}.

\begin{figure*}[!th]
    \centering
    \begin{subfigure}[t]{0.42\textwidth}
    \caption{LinkedIn Example}
    \addtocounter{subfigure}{-1}	 
    \centering
        \begin{subfigure}[t]{0.45\textwidth}
                 \renewcommand\thesubfigure{\alph{subfigure}-1}
        		\includegraphics[width=\linewidth]{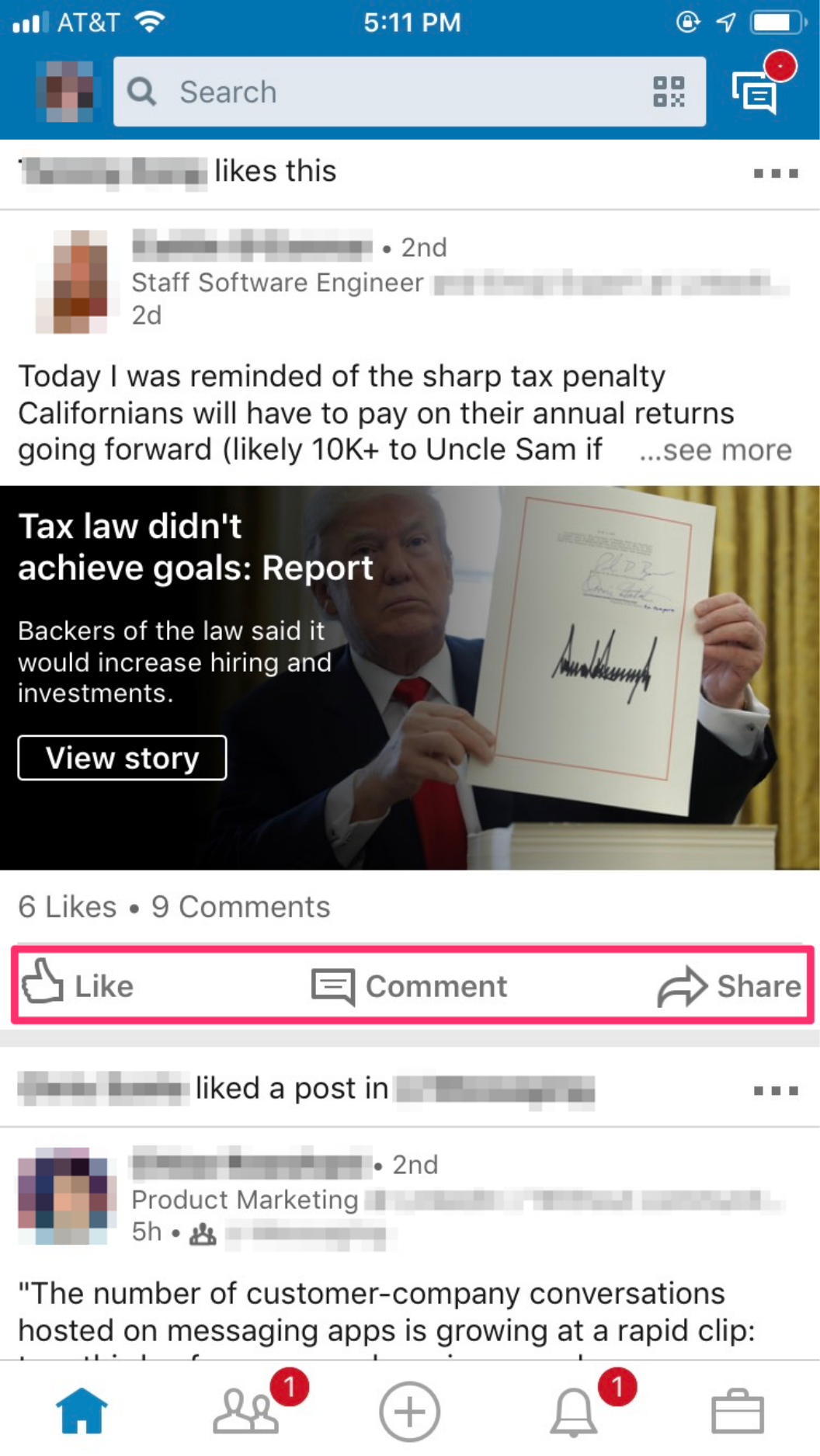}
        		\caption{Mobile Newsfeed}
        		\label{fig:LinkedInFeed}
    	\end{subfigure}%
	\hspace{0.05\textwidth} 
    	\begin{subfigure}[t]{0.45\textwidth}
        	\centering
	        \addtocounter{subfigure}{-1}
	        \renewcommand\thesubfigure{\alph{subfigure}-2}
       		\includegraphics[width=\linewidth]{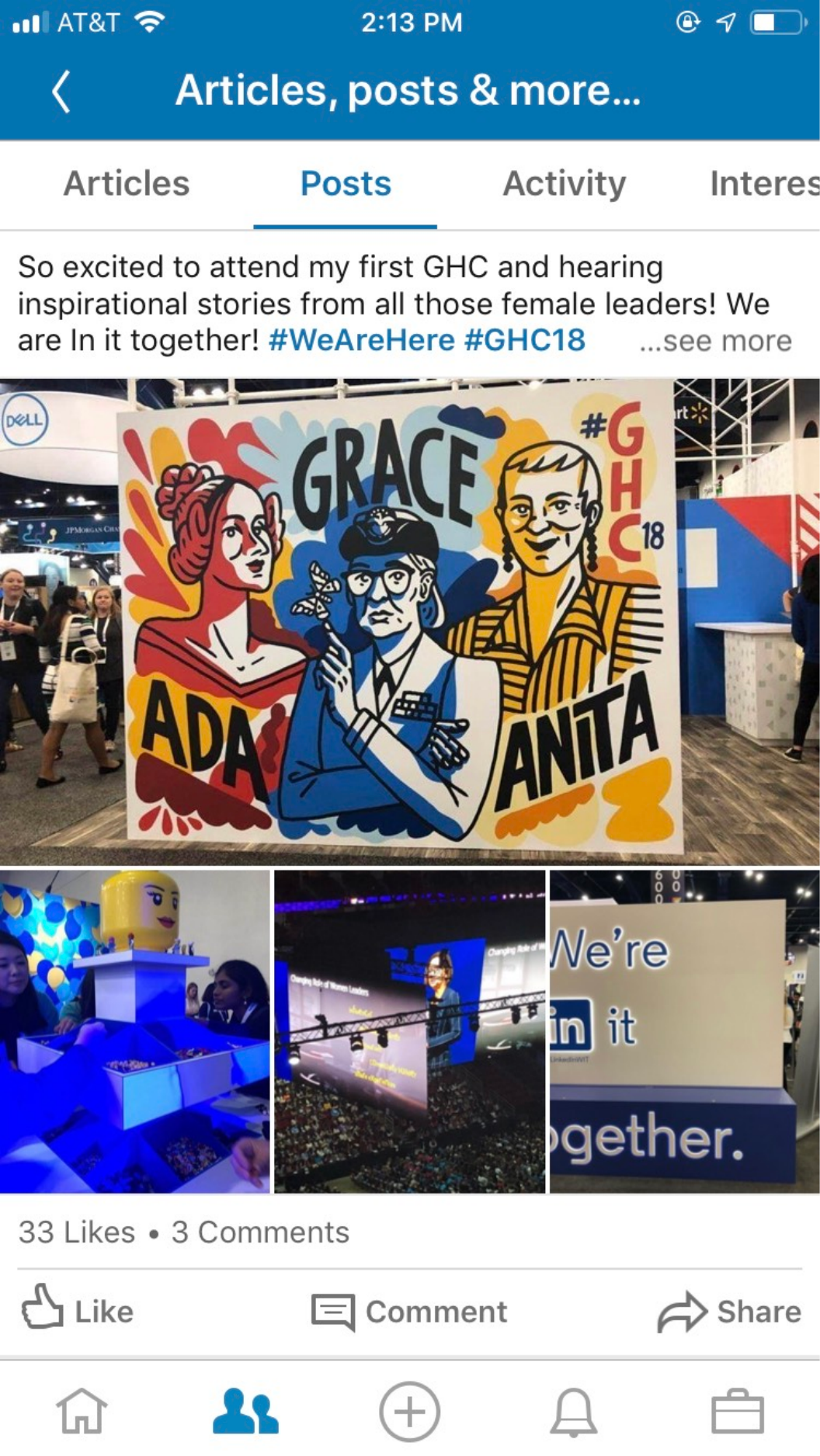}
 		\caption{Creator Activity Tab}
  		\label{fig:LinkedInActivityTab}
   	 \end{subfigure}
	
    	\begin{subfigure}[t]{0.8\textwidth}
	\vspace{0.05\textwidth}
        	\centering
	        \addtocounter{subfigure}{-1}
	        \renewcommand\thesubfigure{\alph{subfigure}-3}
       		\includegraphics[width=\linewidth]{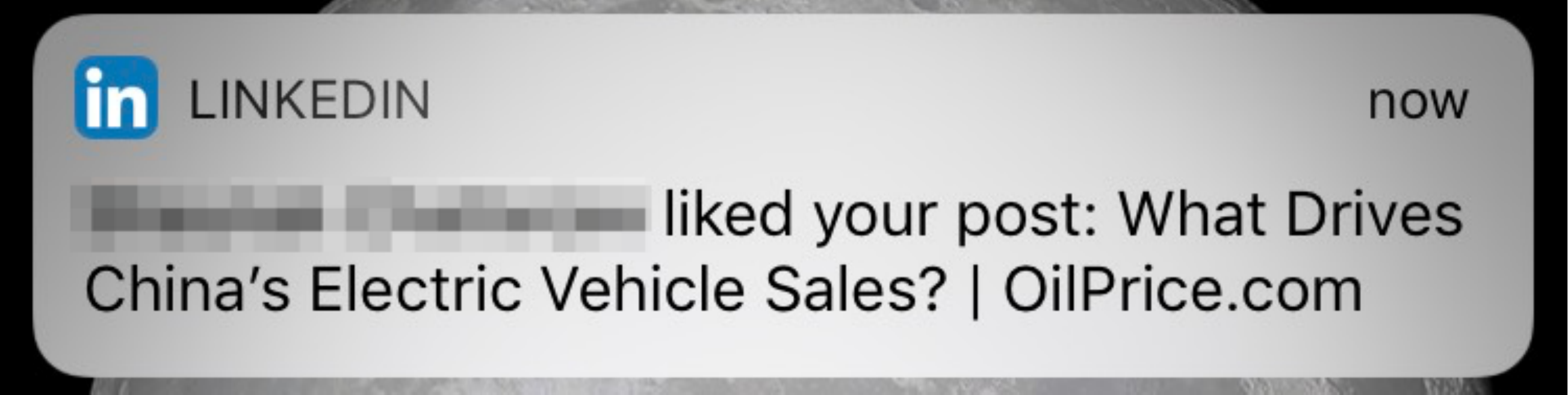}
 		\caption{Creator Notification on Feedback}
  		\label{fig:LinkedInPush}
   	 \end{subfigure}
   \end{subfigure}  
\hfill
    \begin{subfigure}[t]{0.42\textwidth}
    \caption{Instagram Example}
    \addtocounter{subfigure}{-1}	 
    \centering
        \begin{subfigure}[t]{0.45\textwidth}
                 \renewcommand\thesubfigure{\alph{subfigure}-1}
        		\includegraphics[width=\linewidth]{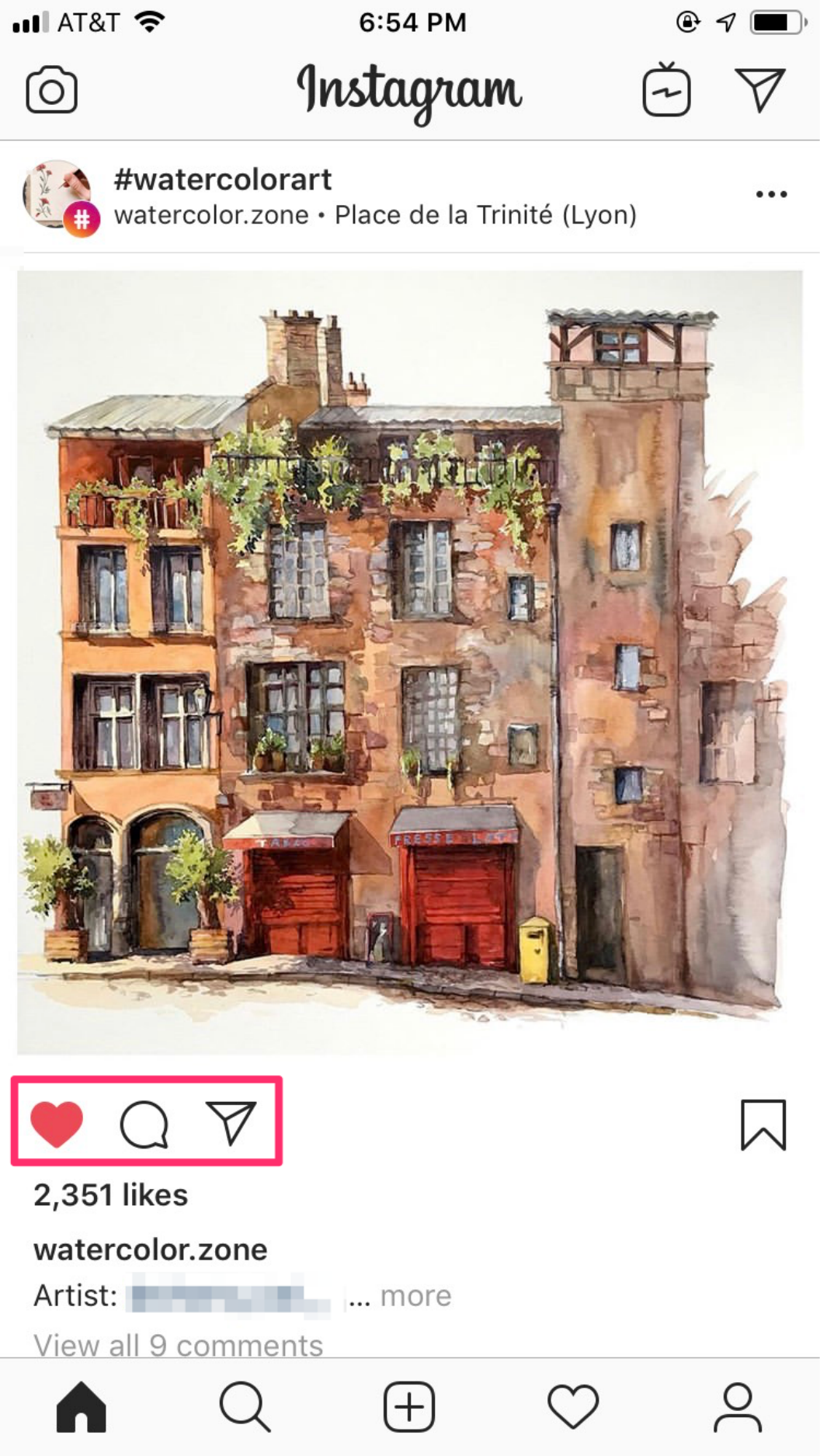}
  		\caption{Mobile Newsfeed}
  		\label{fig:InsFeed}
    	\end{subfigure}%
	\hspace{0.05\textwidth} 
    	\begin{subfigure}[t]{0.45\textwidth}
        	\centering
	        \addtocounter{subfigure}{-1}
	        \renewcommand\thesubfigure{\alph{subfigure}-2}
       		\includegraphics[width=\linewidth]{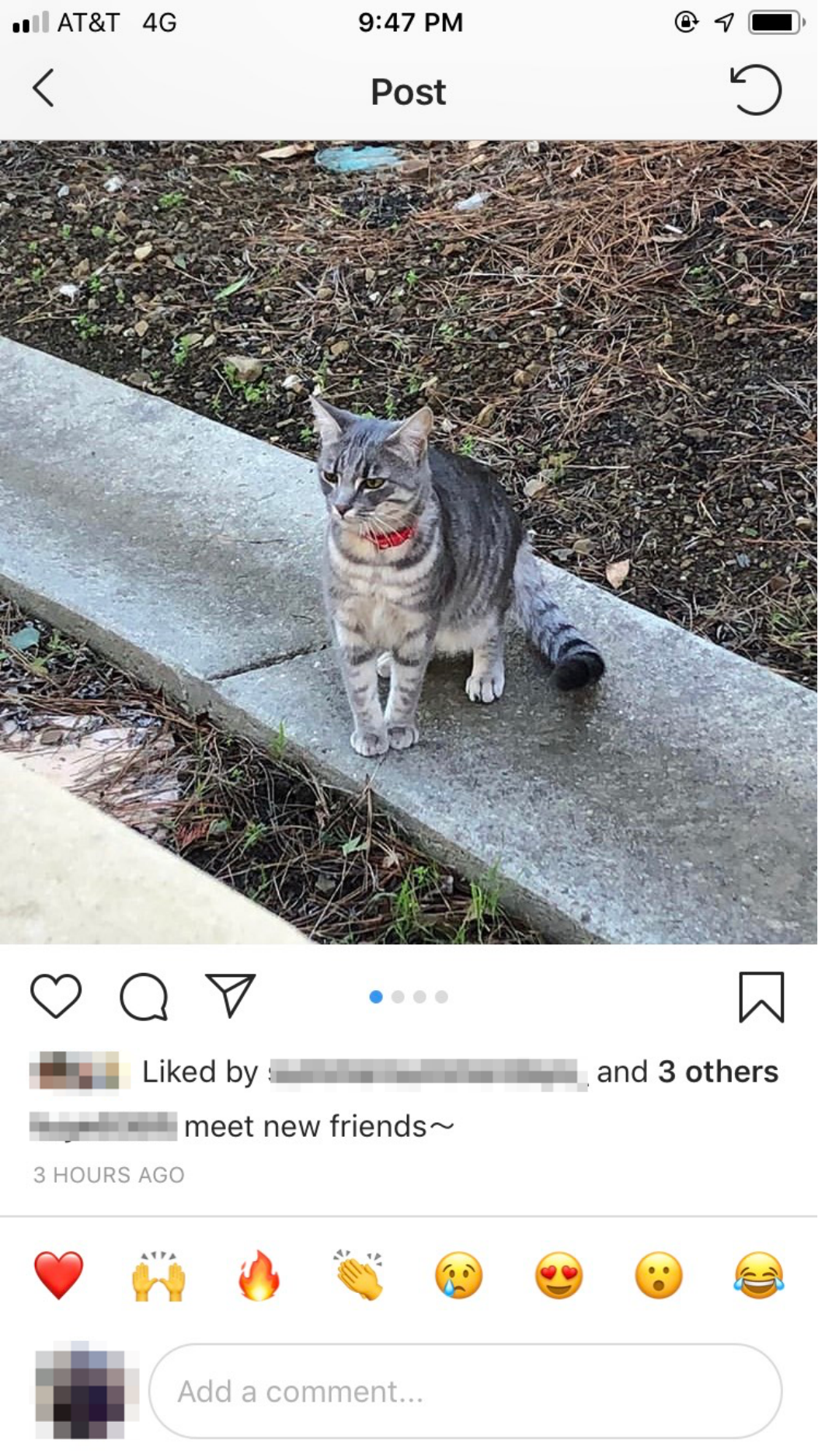}
  		\caption{Creator Activity Tab}
  		\label{fig:InsActivityTab}
   	\end{subfigure}
	
	\begin{subfigure}[t]{0.8\textwidth}
	\vspace{0.05\textwidth}
        	\centering
	        \addtocounter{subfigure}{-1}
	        \renewcommand\thesubfigure{\alph{subfigure}-3}
       		\includegraphics[width=\linewidth]{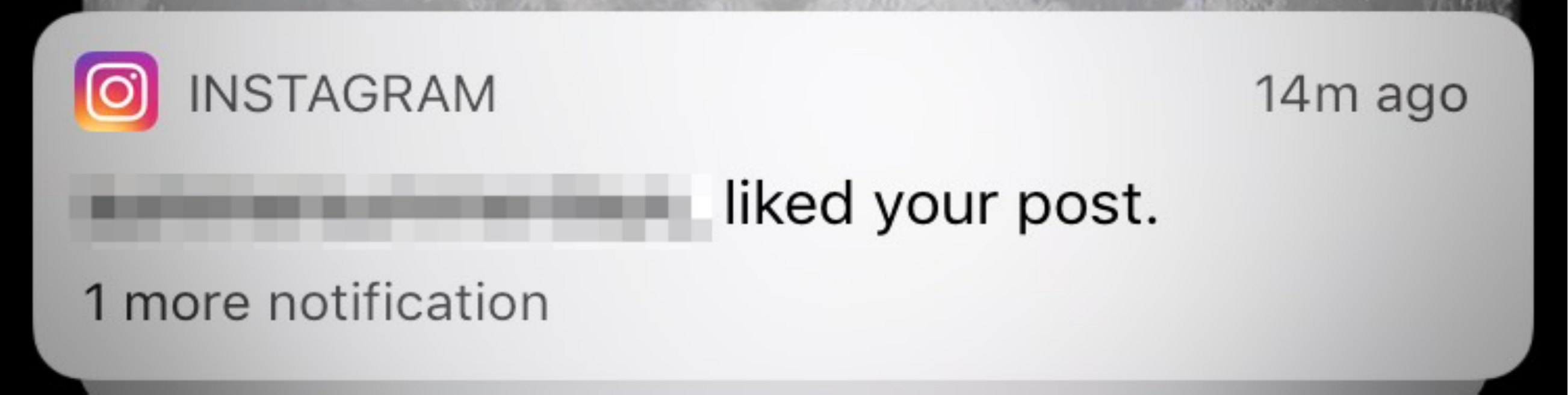}
  		\caption{Creator Notification on Feedback}
  		\label{fig:InsPush}
   	\end{subfigure}
   \end{subfigure} 
    \caption{The different products in the feedback cycle of content creation on LinkedIn and Instagram. Content consumers engage with content via their newsfeed and creators get notified of this feedback via individual notifications, and also via aggregated views on their content activity tab.} 
\end{figure*}

The primary interface for consumers to interact with content and provide feedback to creators is a newsfeed, instances of which are shown in Figures ~\ref{fig:LinkedInFeed} (LinkedIn mobile app) and ~\ref{fig:InsFeed} (Instagram mobile app). Users can also use this interface (e.g., the post button on LinkedIn), or an adjacent screen (as on Instagram) to create and post content. Feedback is received and first presented to creators through notifications as shown in Figures ~\ref{fig:LinkedInPush} and ~\ref{fig:InsPush}, and creators can also get an aggregated view of the feedback and pursue subsequent conversations (e.g., replying to comments) from the self activity dashboard page (examples from LinkedIn and Instagram shown in Figures ~\ref{fig:LinkedInActivityTab} and ~\ref{fig:InsActivityTab}). Other social media platforms like Facebook and WeChat have similar product mechanisms for creators and consumers. Based on the feedback loop, it makes most sense to change the ranking on feed (the primary consumer interface) to provide more effective feedback to creators.

\begin{figure}
	\centering
	\includegraphics[width=\linewidth]{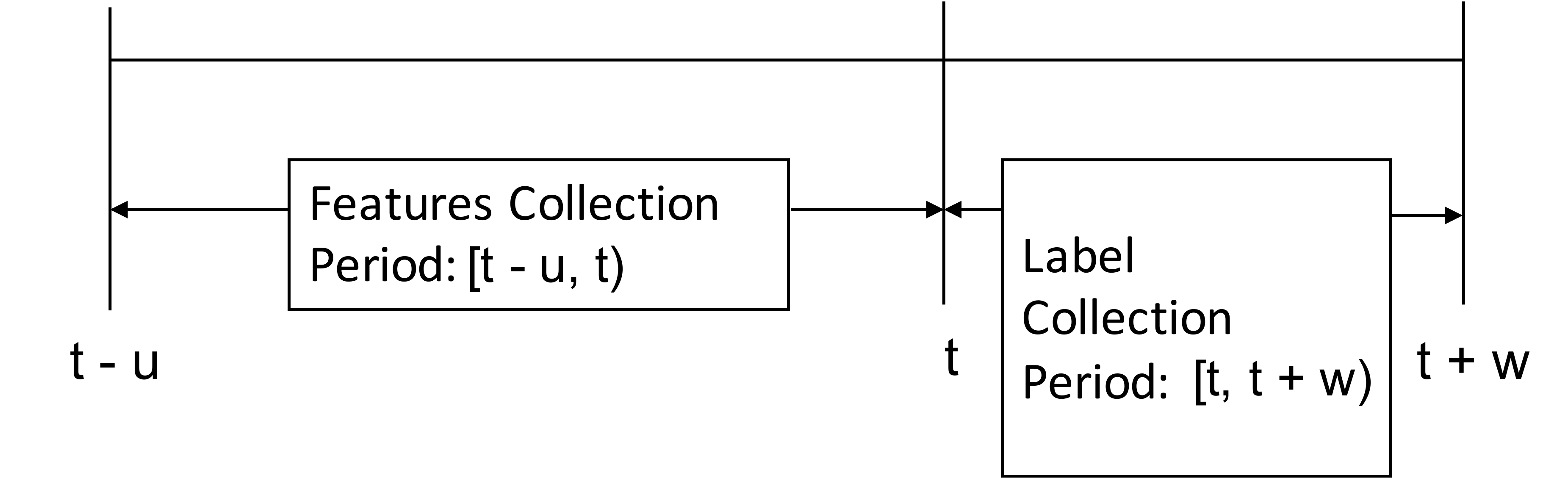}
	\caption{Data collection timeline}
	\label{fig:training}
\end{figure}

We now define the problem formally as follows. We use $i$ as an index over users. Let $t$ denote a discrete time point, $u$ and $w$ represent time intervals, $Y_{i}$ be the number of content pieces user $i$ created during the time window $[t, t+w]$  and $X_{i}$ be the user features for user $i$ computed during the time window $[t-u, t)$. Figure ~\ref{fig:training} illustrates the timeline. We then estimate the creator side utility $V_{i}$ with feedback sensitivity utilities based on the prediction of probability of creating, $P(Y_{i} > 0 \mid X_{i})$ (details will be illustrated in Section \ref{sec:utilityEstimations}). The features, $X_{i}$, comprise of:

\begin{itemize}
    \item $a_{i}$ is the number of feedback items that user $i$ received in the time window $[t-u, t)$. 
    \item $S_{i}$ is the set of other static features of user i at time $t$.
    \item $a_{i} \times S_{i}$ are the interaction terms between the feedback features and the static user features. For instance ``feedback in last 7 days'' $\times$ ``member locale'' of user $i$, which would help personalize the estimation of the effect of feedback.
\end{itemize}

Prior work in optimizing creator side utility at LinkedIn introduced a heuristic model ~\cite{bonnie2018}. Our goal is to improve this baseline heuristic with a model-based feedback sensitivity utility.

\section{Modeling Creation Behavior and Utility Estimations}
\label{sec:utilityEstimations}

We now describe our future creation prediction model, and how that is used to come up with the feedback sensitivity estimation. Some of our modeling choices are influenced by member engagement patterns on LinkedIn, where different choices may be more pertinent in a non-LinkedIn context. To make the narrative accurate yet generally informative, we will clearly delineate those instances and also present the abstract learnings which are more broadly applicable.


\subsection{Response Prediction Models}
As we plan to serve the utilities in large scale online recommendation platforms where the latency tolerance is low, we focus on developing scalable linear or non-linear models to quantify how effective feedback can motivate a content creator to create again in the future on LinkedIn. 

\subsubsection{pCreate Models}
The objective of the "pCreate'' model is to predict the probability of user $i$ creating at least one content piece during the time window $[t, t+w]$ given all the features, i.e., $P(Y_{i} > 0\mid X_{i})$. Instead of predicting the number of content pieces created, we use the binarized version of the response because the number of creations for frequent creators is quite noisy, and the binarization retains most of the information since only a small fraction of all creators are constant creators. In a general setting, it may be more useful to formulate this as a regression problem with $Y_{i}$ as the response, and an appropriate link function depending on the distribution of $Y_{i}$.

\paragraph{Linear Logistic Model}
We use a linear logistic function to model the probability of creation as follows:

\begin{equation}
\begin{aligned}
\centering
p_{i} & = P(Y_{i} > 0 \mid X_{i}) \\
& = \frac{1} {1+ e^{- (\mu + \gamma^T S_{i} + \lambda a_{i} + {\beta}^T \{a_{i} \times S_{i}\})}} ,
\end{aligned}
\label{eq:linearmodel}
\end{equation}
where $\mu$ is the global intercept, $\lambda$, $\beta^T$, and $\gamma^T$ are the coefficients for $a_{i}$, $S_{i}$, and $a_{i} \times S_{i}$.

\begin{figure}
  \centering
  \includegraphics[width=1\linewidth]{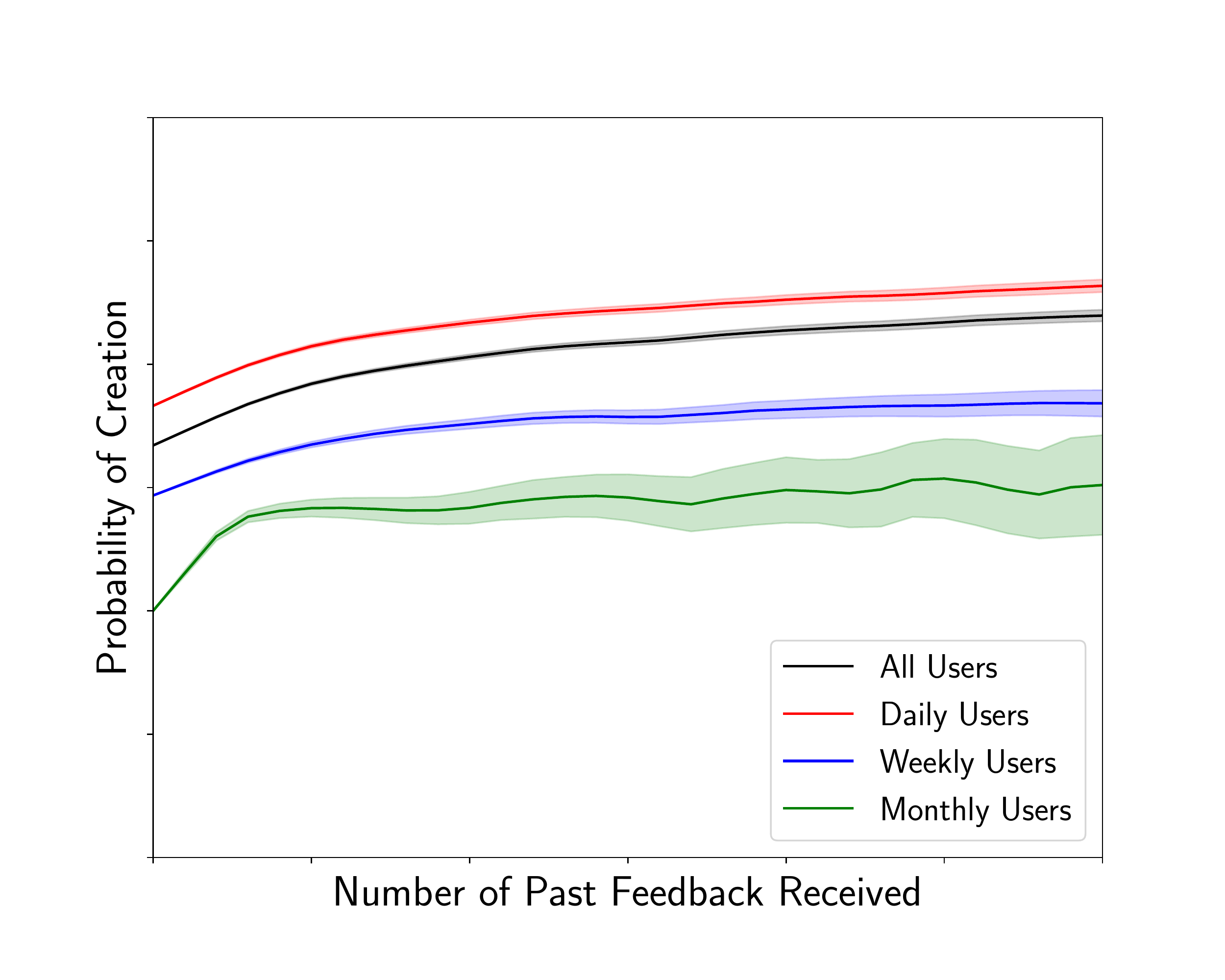}
  \caption{Probability of Creation (within $95\%$ confidence interval) vs Feedback}
  \label{fig:feedback}
\end{figure}

The relationship between feedback and a user's future creation is shown in Figure ~\ref{fig:feedback}. The plot demonstrates three key things: 
\begin{enumerate*}
\item The incremental value from feedback tends to diminish with increasing feedback amount;
\item The impact of feedback on creation varies with certain member features like activity level; and 
\item The non-linearity between probability of creation and feedback can not be captured by the sigmoid function (the logit of the response would still be non-linear with feedback).
\end{enumerate*}

To address the non-linearity between $logit(P(Y_{i})>0)$ and $a_{i}$, we bucketized the numerical feedback feature $a_{i}$ into a $k$-level categorical feature (the exact bucketization points are guided by the above correlation analysis) to allow for flexibility in the estimation of the feedback effect. We also include feedback interaction features $\{ a_{i} \times S_{i} \}$ in the linear model to better distinguish among member cohorts on how feedback affects future creation. 


\paragraph{Tree Model}
\label{sec:nonlinearModel}
The non-linearity between features and response can also be handled by a tree model. To this end, we used an ensemble model of decision trees via the XGBoost library ~\cite{Chen:2016:XST:2939672.2939785}. This approach no longer requires the bucketization of the feedback feature $a_i$, or the explicit use of interaction features $\{a_i \times S_i\}$.

%

\subsection{Estimation of Feedback Sensitivity Utility}
\label{sec:estimationOfFeedbackSensitivityUtility}
We define the ``feedback sensitivity utility'' as a \textit{delta probability of creation given an increase of feedback}. The delta score can be estimated as follows:

\begin{enumerate}
\item Fetch all features $X_{i}$ for member $i$.
\item Derive the updated feedback feature vector given the user $i$ receives one incremental unit of feedback.
\item Calculate the utility using estimated probability of content creation $\hat{P}(Y_{i} > 0)$ as
\begin{equation}
\begin{aligned}
\centering
\delta p_{i} &= \Delta \hat{P}(Y_{i} > 0 \mid X_{i}) \\
& =  \hat{P}(Y_{i} > 0 \mid (a_{i} + \Delta a_{i}), X_{i}) \\
& \quad - \hat{P}(Y_{i} > 0 \mid a_{i}, X_{i})
\end{aligned}
\label{eq:deltaprob}
\end{equation}
\end{enumerate}

\subsubsection{Fit Utility to Exponential Decay Form}
Data sparsity is a major challenge in modeling user content creation behavior, especially for less active users and creators. The sparsity can make the delta estimates of such users quite noisy, hence we use a parameterized approach to smooth these estimates. Based on the observation in Figure ~\ref{fig:feedback}, we use an exponential decay form. The utility function is then fitted using the following process.


\begin{enumerate}
\item Let $V$ be a $K \times 2$ matrix as shown in Equation ~\ref{eq:V}, where $v_{k}$ represents a list of possible values of the ``feedback received'' feature, and each value is a representative of the feedback value interval (we use the minimum value of the interval) it represents:
\begin{equation}
\label{eq:V}
V =   
\begin{bmatrix}
    1&v_{1} \\
    1&v_{2} \\
    \vdots&\vdots\\
    1&v_{K}\\
\end{bmatrix}
\end{equation}
\item Estimate the delta effect of feedback as the change in creation probability as we transition from one feedback level to the next higher one. $v_{0}$ is the initial feedback level of $0$ feedback.
\begin{equation}
 \delta \hat{p}_{i,k} =  \frac{\hat{P}(Y_{i,k}>0\mid a_{i} = v_{k},X_i) - \hat{P}(Y_{i,k} > 0\mid a_{i}=v_{k-1},X_i)}{v_k - v_{k-1}}
\end{equation}
\item Let $\delta \hat{P}_{i}$ be a $K \times 1$ vector of the level-specific feedback sensitivity of user $i$:
\begin{equation}
\label{eq:P}
\delta \hat{P}_{i}=   
\begin{bmatrix}
    \delta \hat{p}_{i,1} \\
    \vdots\\
    \delta \hat{p}_{i,K}\\
\end{bmatrix}
\end{equation}
\item Now to fit the user-specific curve, let $\tau_i$ denote the exponential decay factor, and $b_i$ denote an offset. In order to fit these two parameters, we have $K$ data points:
\begin{equation}
\delta \hat{P}_{i} = e^{V [ b_{i} \; \tau_{i} ]^T + \epsilon}
\label{eq:expform1}
\end{equation}
\item Taking logarithm on both sides of the Equation ~\ref{eq:expform1}, we have:
\begin{equation}
\log(\delta \hat{P}_{i}) = V [b_{i}\; \tau_{i}]^T + \epsilon
\label{eq:expform2}
\end{equation}
\item We can solve for $\tau_{i}$ and $b_{i}$ for each user $i$ as Equation ~\ref{eq:tau_intercept} shown.
\begin{equation}
\label{eq:tau_intercept}
\begin{bmatrix}
\hat{b}_{i}\\
\hat{\tau}_{i} 
\end{bmatrix}
= ({V}^TV)^{-1}{V}^T \log(\delta \hat{P}_{i})
\end{equation}
\end{enumerate}

When we train trees using XGBoost, we use the actual, instead of bucketized, feedback value as feature. However, when fitting $\tau_{i}$ and $b_{i}$, we currently use the same feedback intervals as in Logistic model. Alternately, we could have the tree model learn $K$ intervals, and generate a system of $K$ data points per user, sampling $1$ data point per learnt interval.

\subsection{Data Collection}
\label{sec:dataCollection}
Clean and high quality training data is critical to building a good model. We sample from one month of ~100M active users' data at LinkedIn, and track their creation behavior. We include three major categories of features ($X_{i}$) as the following:
\begin{itemize}
\item $a_{i}$: Feedback features, the intervention feature.
\item $S_{i}$: User profile features, such as country, preferred language, job title, etc. Also, static features from the user's network (e.g., number of connections, followers and followees).
\item $X_{i} \setminus a_{i}$: Activity features including past visits, feed consumption features, likes, shares, comments, etc.
\end{itemize}

The data collection process is the same as described previously (and shown in Figure ~\ref{fig:training}). The response interval $w$ can be defined as $1$ day, $1$ week, or longer --- we use \textit{$1$ week}. We compute the dynamic (activity based) features using one month data. Hence, the interval $u$ is $1$ month. We collect the static (profile/network based) features as a snapshot of the user's status at time $t$. 

\subsection{Offline Evaluation Analysis}
\label{sec:offlineanalysis}


This section presents the offline performance of the logistic regression model and the XGBoost model to predict content creation. The collected data is partitioned randomly into training, validation and testing datasets. The validation dataset is used for hyper-parameter tuning (i.e., the regularization parameters in both logistic and tree models). Besides the model performance on all users, we also look at the results for specific member cohorts, segmented by member activity levels and contribution levels. Member activity levels represent the frequency of LinkedIn visits, while contribution levels represent the frequency of generating or distributing (e.g., share, like, comment, message) content contributions. \textit{Contributions are a super-set of creation, and the relaxed definition makes the least active cohorts less sparse}. The segmented view allows us to understand the current model performance better and identify areas of improvement.

\subsubsection{Logistic regression model performance}

\begin{figure*}[!th]
   \centering
    \begin{subfigure}[t]{0.33\textwidth}
    \caption{Model Performance}
    \addtocounter{subfigure}{-1}	 
    \centering
        \begin{subfigure}[t]{1.1\textwidth}
                 \renewcommand\thesubfigure{\alph{subfigure}-1}
        		\includegraphics[width=\linewidth]{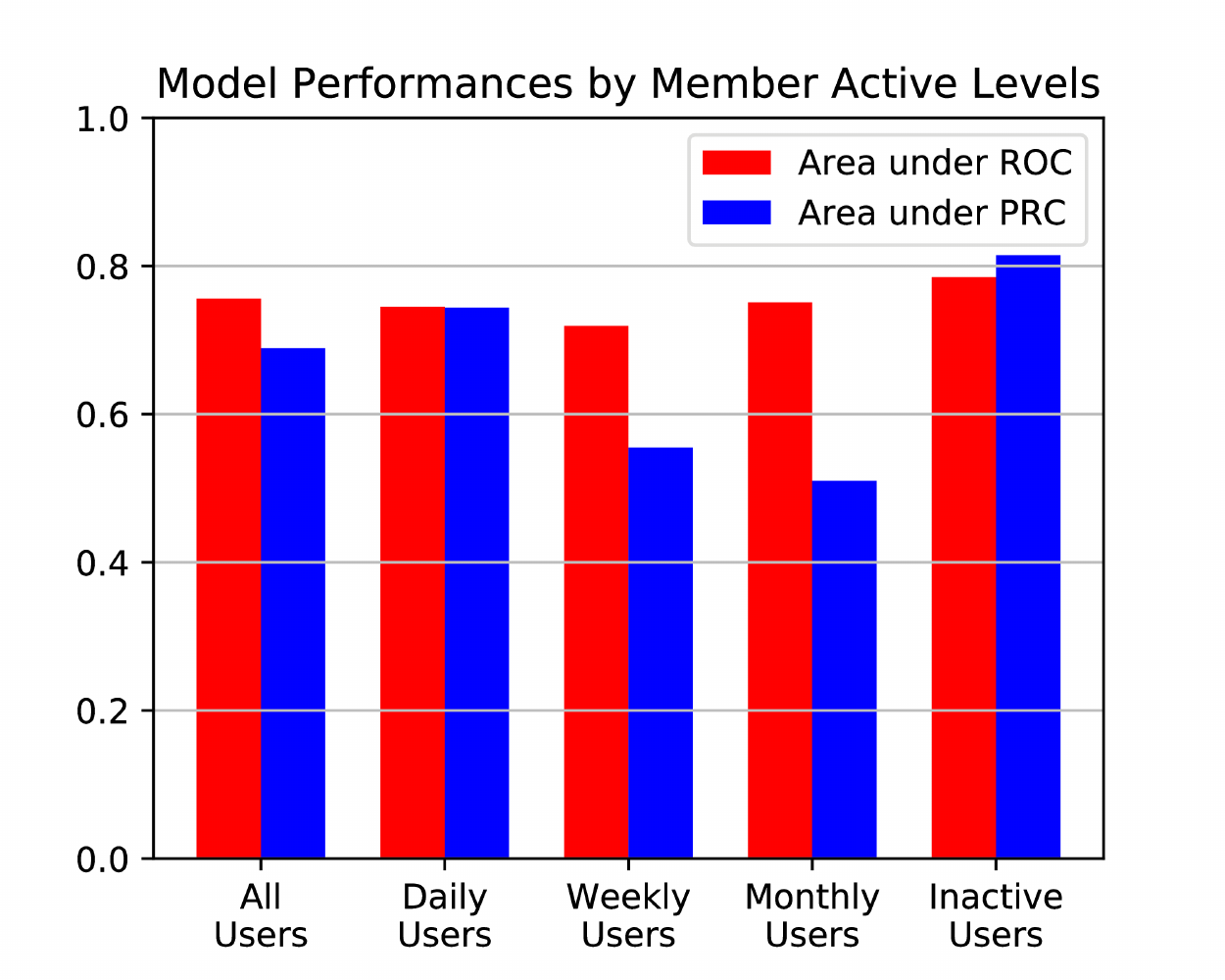}
        		\caption{Segmented by Activity Levels}
        		\label{fig:Model_performance_mlc}
    	\end{subfigure}
	\vspace{0.05\textwidth} 
	
    	\begin{subfigure}[t]{1.1\textwidth}
        	\centering
	        \addtocounter{subfigure}{-1}
	        \renewcommand\thesubfigure{\alph{subfigure}-2}
       		\includegraphics[width=\linewidth]{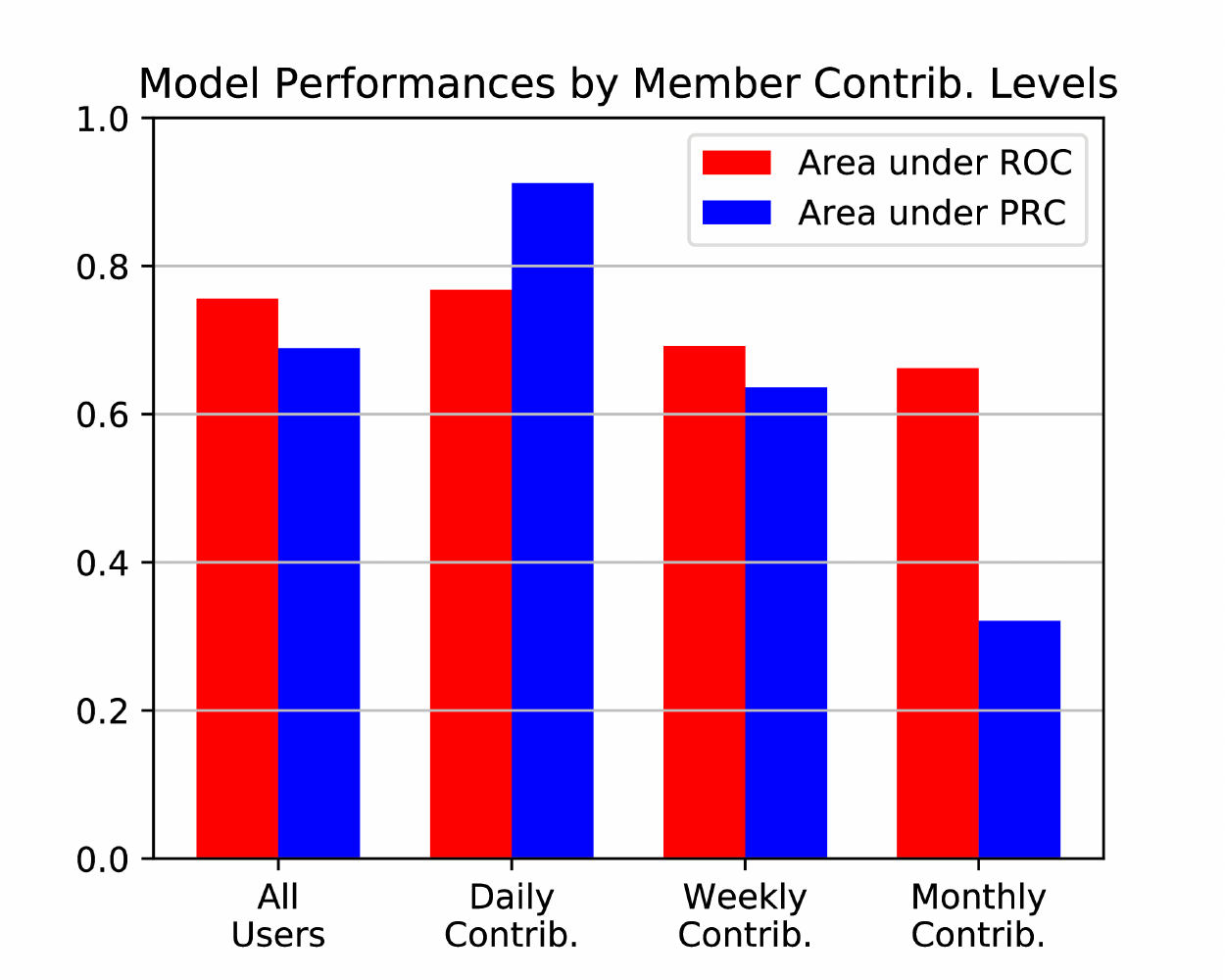}
 		\caption{Segmented by Contrib. Levels}
  		\label{fig:Model_performance_clc}
   	 \end{subfigure}
   \label{fig:Model_performance}
   \end{subfigure}  
 \hfill
    \begin{subfigure}[t]{0.33\textwidth}
    \caption{Feedback Sensitivity}
    \addtocounter{subfigure}{-1}	 
    \centering
        \begin{subfigure}[t]{1.1\textwidth}
                 \renewcommand\thesubfigure{\alph{subfigure}-1}
        		\includegraphics[width=\linewidth]{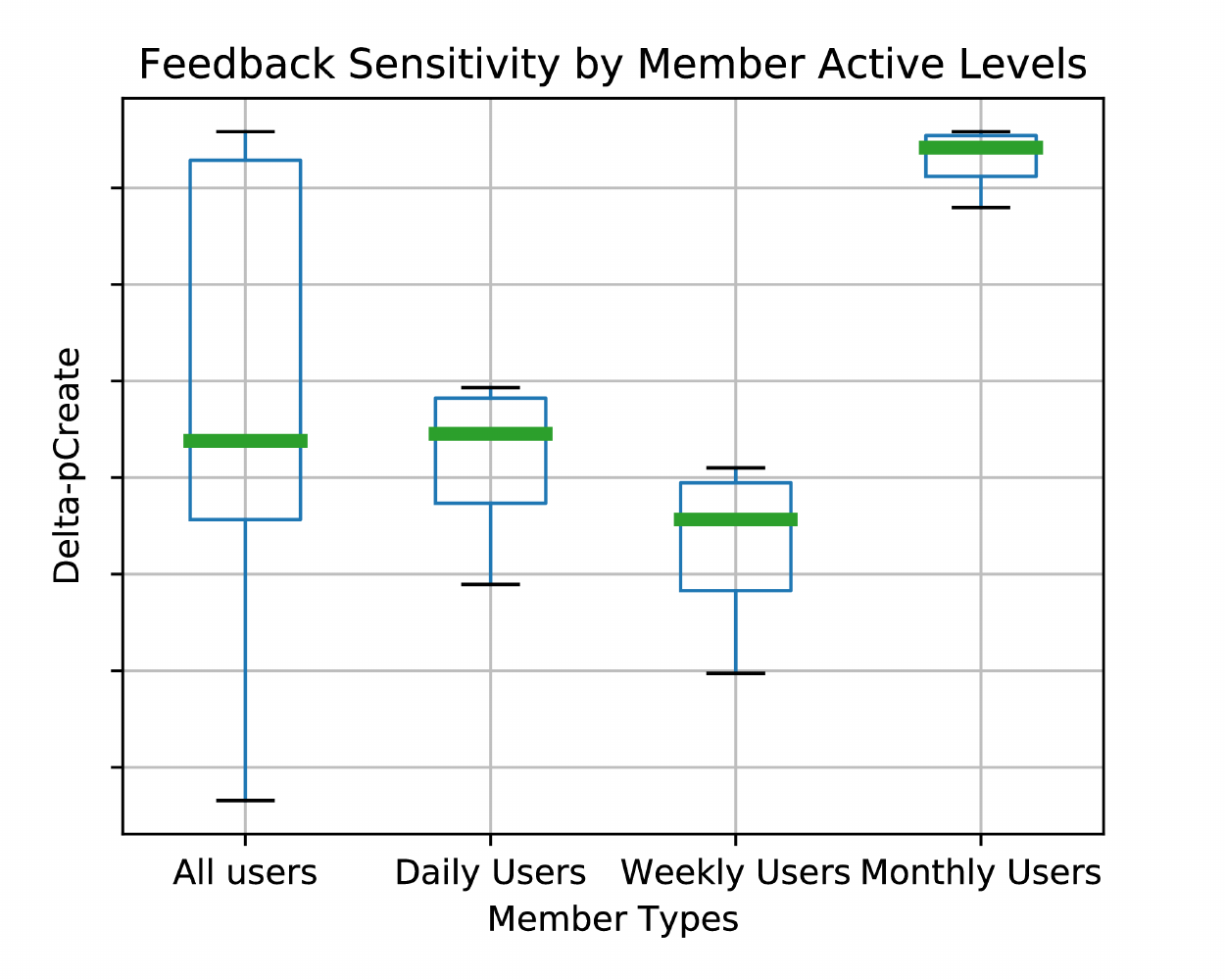}
        		\caption{Segmented by Activity Levels}
        		\label{fig:delta_mlc}
    	\end{subfigure}
	\vspace{0.05\textwidth} 
	
    	\begin{subfigure}[t]{1.1\textwidth}
        	\centering
	        \addtocounter{subfigure}{-1}
	        \renewcommand\thesubfigure{\alph{subfigure}-2}
       		\includegraphics[width=\linewidth]{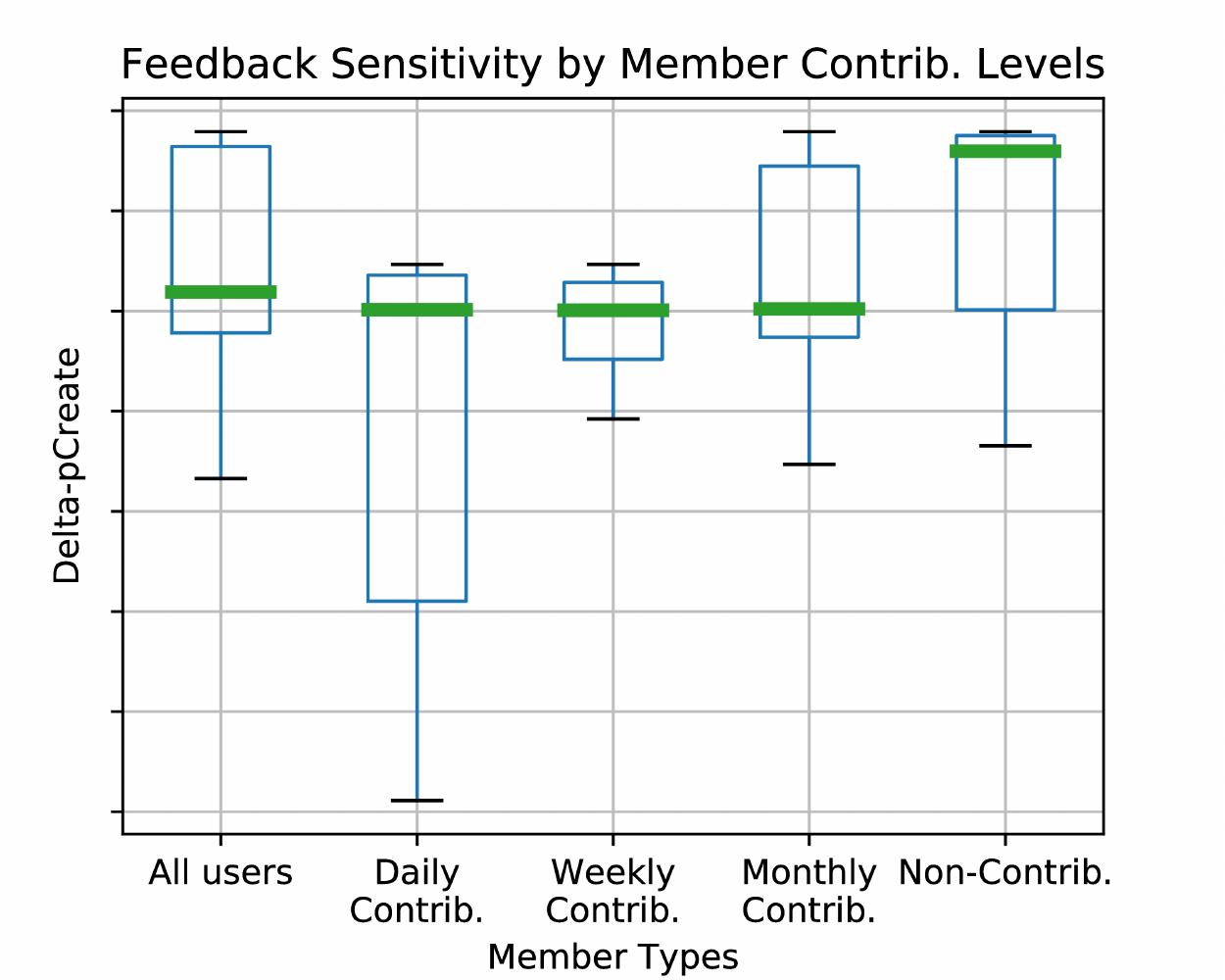}
 		\caption{Segmented by Contrib. Levels}
  		\label{fig:delta_clc}
   	 \end{subfigure}
   \label{fig:delta}
   \end{subfigure} 
\hfill
    \begin{subfigure}[t]{0.33\textwidth}
    \caption{Feedback Sensitivity (Expon. Decay Model)}
    \addtocounter{subfigure}{-1}	 
    \centering
        \begin{subfigure}[t]{1.1\textwidth}
                 \renewcommand\thesubfigure{\alph{subfigure}-1}
        		\includegraphics[width=\linewidth]{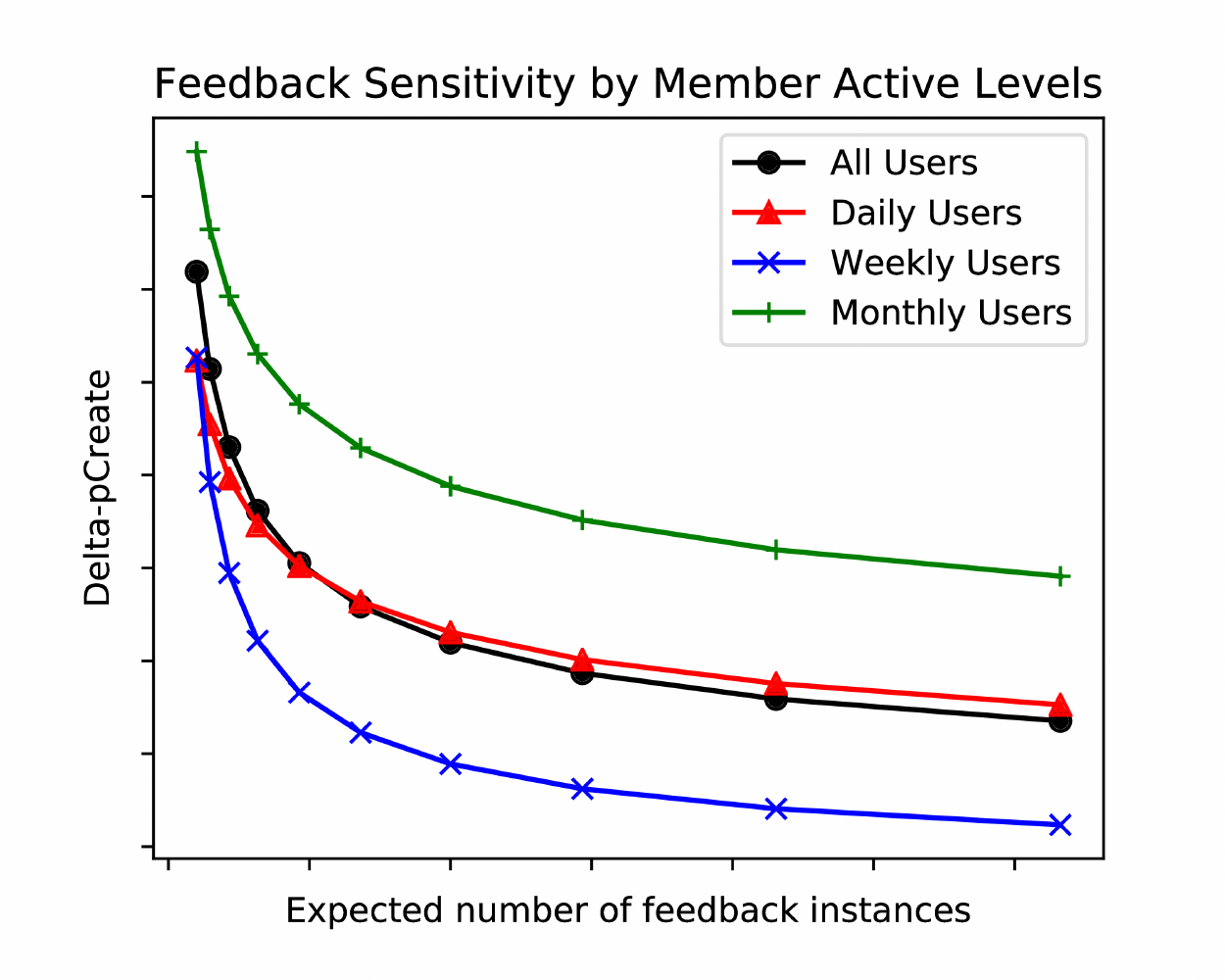}
        		\caption{Segmented by Activity Levels}
        		\label{fig:tau_mlc}
    	\end{subfigure}
	\vspace{0.05\textwidth} 
	
    	\begin{subfigure}[t]{1.1\textwidth}
        	\centering
	        \addtocounter{subfigure}{-1}
	        \renewcommand\thesubfigure{\alph{subfigure}-2}
       		\includegraphics[width=\linewidth]{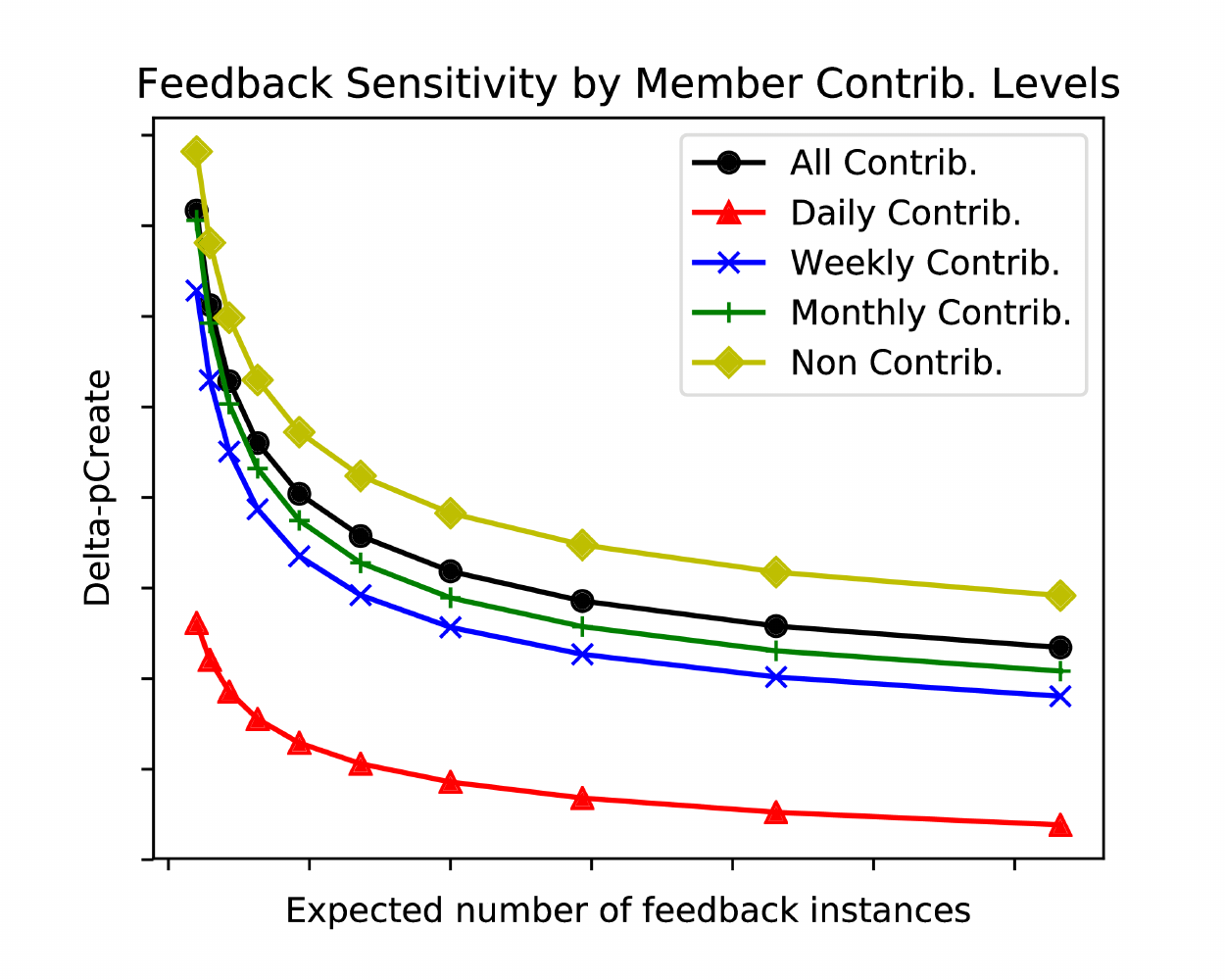}
 		\caption{Segmented by Contrib. Levels}
  		\label{fig:tau_clc}
   	 \end{subfigure}
   \label{fig:tau}
   \end{subfigure} 

    \caption{Offline performance of the logistic regression model, and feedback sensitivity, by member cohorts} 
\end{figure*}

Figure ~\ref{fig:Model_performance} shows the performance of the logistic regression model in terms of the area under ROC (AUROC) and the area under Precision-Recall curve (AUPRC). The model achieved $0.756$ in AUROC and $0.689$ in AUPRC across all users. When sliced by user activity levels (Figure ~\ref{fig:Model_performance_mlc}), the model has the best performance for inactive users, followed by daily users. Weekly users and monthly users' content creation behavior is more difficult to predict. Slicing by user contribution levels (Figure ~\ref{fig:Model_performance_clc}), daily contributors achieved the best model performance, followed by weekly and then monthly contributors. Both the segmented results indicate that the model performance improves with higher activity and creation levels. This is likely because users who visit LinkedIn frequently and/or contribute more frequently provide more data for training and their activities are more structured. The high accuracy for Inactive users has very low confidence (and hence is noisy), as there is very little data for that segment. Since they also account for less than $1\%$ of all data, we will exclude them from the rest of the discussion.

\subsubsection{Feedback sensitivity}
As explained in Section ~\ref{sec:estimationOfFeedbackSensitivityUtility}, we utilized the model results in either the feedback sensitivity form ($\delta \hat{p}_{i,k}$), or the exponential decay form ($\hat{\tau}_{i}$ and $\hat{b}_{i}$). The results of these two forms, again segmented by member activity levels and contribution levels, are shown in Figure ~\ref{fig:delta} and Figure ~\ref{fig:tau}. The segmented analyses for feedback sensitivity identifies the most and least feedback sensitive cohorts.

As can be seen in Figure ~\ref{fig:delta_mlc} and Figure ~\ref{fig:delta_clc}, feedback sensitivity level is more diverse and separable when segmented by activity levels as compared to contribution levels. Monthly users are significantly more sensitive to feedback than daily and weekly users, and daily users has slightly higher feedback sensitivity than weekly users. This is likely because Monthly users receive less feedback than more active Daily and Weekly users, and hence additional feedback has higher value to them. This also aligns with the analysis shown in Figure ~\ref{fig:feedback}. Among the various contribution levels, daily, weekly and monthly contributors have comparable median feedback sensitivities, with good overlap in the range as well. Non-contributors (who have not contributed in the past 4 weeks) have significantly higher median feedback sensitivity. Therefore, infrequent contributors who still actively visit LinkedIn (at least once a month) are quite sensitive to feedback.

Figure ~\ref{fig:tau} shows the feedback sensitivity after fitting the utility in the exponential decay form. The x-axis represents the expected number of feedback instances and y-axis represents feedback sensitivity of receiving one more feedback instance. The fitted feedback sensitivities under different cohorts are consistent (in terms of relative orders) with the results shown in Figure ~\ref{fig:delta}. The use of fitted feedback sensitivity is preferred when the number of cohorts is large as the data of some cohorts could be sparse. Although different member types have slightly different rates of decay, the order of sensitivity does not change as the expected feedback instance increases. Therefore, member type is a more important factor when the expected feedback level is similar. 

\subsubsection{Tree model performance}
Table ~\ref{tab:XGBoostPerformance} shows the performance of the XGBoost model and its comparison to the logistic regression model for different user activity levels. The same set of data was used for both model performance analyses. However, the XGBoost model did not use interaction features, and the feedback feature is not bucketized, as tree models inherently learn nonlinear relationships and allow the formation of discriminative interaction features. The XGBoost model performed better for all member cohorts. Overall, the XGBoost model achieved $+1.9\%$ in AUROC and $+2.5\%$ in AUPRC as compared to the logistic regression model. The improvement is more significant for weekly users and monthly users, especially in AUPRC (\textit{e.g.}, $+4.8\%$ for weekly users and $+16.3\%$ for monthly users in AUPRC). As the weekly and monthly users are more sensitive to feedback (illustrated in Figure ~\ref{fig:delta}), improving model accuracy for these members is most beneficial. The improvement from the non-linear model (compared to the linear model) is expected. We were more interested in quantifying the extent of the improvement.

\begin{table}[h]
	\centering
	\begin{tabular}{l c ccc} 
		\hline\hline 
		Member Types & Metrics  & Logistic & XGBoost& Gain
		\\ [0.5ex]
		\hline 
		&AUROC & $0.756$ & $0.770$ & $+1.9\%$  \\[-1ex]
		\raisebox{1.5ex}{All Users} 
		&AUPRC & 0.690 & $0.709$ & $+2.6\%$  \\[1ex]
		&AUROC & $0.744$ & $0.757$ & $+1.7\%$  \\[-1ex]
		\raisebox{1.5ex}{Daily Users} 
		&AUPRC &0.742 & $0.757$ & $+2.0\%$  \\[1ex]
		&AUROC & $0.718$ & $0.736$ & $+2.5\%$  \\[-1ex]
		\raisebox{1.5ex}{Weekly Users} 
		&AUPRC & 0.557 & $0.584$ & $+4.8\%$  \\[1ex]
		&AUROC & $0.751$ & $0.790$ & $+5.2\%$  \\[-1ex]
		\raisebox{1.5ex}{Monthly Users} 
		&AUPRC & 0.502 & $0.584$ & $+16.3\%$  \\[1ex]
		&AUROC & $0.781$ & $0.829$ & $+6.1\%$  \\[-1ex]
		\raisebox{1.5ex}{Inactive Users} 
		&AUPRC & 0.820 & $0.864$ & $+6.4\%$  \\[1ex]
		
		\hline 
	\end{tabular}
	\caption{Offline performance of prediction models} 
	\label{tab:XGBoostPerformance}
\end{table}

\section{Feed Recommendation Application}
\label{sec:feedApplication}
We now discuss how to use the "pCreate'' model-based feedback sensitivity to modify feed ranking and incentivize content creators. We introduced two forms of feedback sensitivity related utilities: 
\begin{enumerate*}
\item
$\delta \hat{p}_{i}$ as a direct proxy of creator utility represents the incremental probability of a specific creator $i$ generating some content if he or she receives higher amount of feedback;
\item $\hat{\tau}_{i}$ and $\hat{b}_{i}$ represents the decay rate and intercept of feedback value to creation assuming an exponential decay form of the incremental feedback value. 
\end{enumerate*}

We modify the feed ranking score formulation to incorporate the utilities (as described in Section ~\ref{sec:modifyfeed}) and launched online A/B tests to compare the pCreate model-based feedback utilities with the baseline model (results in Section ~\ref{sec:onlineTest}).

\subsection{Modifying feed ranking}
\label{sec:modifyfeed} 
If an item $k$, whose creator is user $i$, is being shown to a consumer $j$, then its feed score $FeedScore(k,j)$ can be written as:
\begin{equation}
\begin{aligned}
FeedScore(k,j) = &\alpha E[ConsumerUtility(k,j)] \\
& \quad \quad + (1-\alpha) pFeedback(i, j, k) C_{i},
\label{eq:feedscoredelta}
\end{aligned}
\end{equation}
where $C_{i}$ is the creator side utility for $[t, t+w]$, $pFeedback(i, j, k)$ is the probability of consumer $j$ reacting to item $k$ from creator $i$ which gives feedback to creator $i$ and $\alpha$ is a parameter that controls how much we prioritize the consumer utility over the creator utility. Following are the representations of the creator utility, $C_{i}$, in the heuristic approach and the pCreate model.

\subsubsection{Baseline Heuristic Model}
\label{sec:basemodel}
The heuristic model leverages a prediction model on the expectation of feedback ($E(a_{i}$)) assuming an exponential decay function between feedback and the creation utility. The creator side utility is defined formally as
\begin{equation}
\centering
C_{i} = e^{-E(a_{i})}
\label{eq:baseheuristic}
\end{equation}

\subsubsection{pCreate Model}
We represent the creator side utility using the delta feedback effect as:
\begin{equation}
C_{i} = \delta p_{i},
\label{eq:vactorDelta}
\end{equation}
One can also explore models which make $\delta p_{i}$ specific to the consumer $j$ as well, but that is a direction we do not explore more in this work. An alternative is to represent the creator utility assuming the exponential decay of the incremental feedback value as:
\begin{equation}
C_{i} = e^{\tau_{i} E(a_{i}) + b_{i}},
\label{eq:vactorTau}
\end{equation}

We chose the parameterized version of utilities ($\hat{\tau}_{i}$, $\hat{b}_{i}$) due to a few practical advantages over the direct delta effect estimator: 
\begin{itemize}
\item With the creator utility definition in Equation \ref{eq:vactorTau}, we can keep our feedback sensitivity estimation flows offline, and still leverage the real-time estimations of expected feedback, $E(a_{i})$, using the online features.
\item The resultant utility scores are smoother than $\delta \hat{p}_{i}$, especially for the users who are infrequent creators, or who have received very little feedback in the past.
\end{itemize}

\subsection{Online A/B Tests}
\label{sec:onlineTest}

We set up online A/B experiments on both consumers and creators separately to evaluate the performance of our pCreate model against the baseline heuristic. Table \ref{tab:feed_metrics} shows the list of metrics that we are interested in. They represent metrics for consumer engagement as well as creator activity.


\begin{table}[h]
\centering
\begin{tabular}{p{2.5cm}|p{5.2cm}} 
\hline\hline 
Metrics & Descriptions
\\ [0.5ex]
\hline
Contributions & Number of activities that generate or distribute content in the ecosystem (includes the public ones: shares, likes, comments; and the private ones like messages). \\
\hline
(Public/Private) Contributors & Unique number of users who have (public/private) contribution activities.\\
\hline
Contributors with Response & Unique number of contributors who receive a response within a time window.\\
\hline
Retained Creators & Unique number of content creators who are retained from a previous time window.\\
\hline
Feed Viral Actions & Total numbers of viral actions (like, share, comment) on Feed.\\
\hline
Feed Viral Actors & Unique number of users who make viral actions.\\
\hline
Feed Interactions & Total (clicks + viral actions) on Feed.\\
\hline
\end{tabular}
\caption{Metrics of Interest} 
\label{tab:feed_metrics}
\vspace{-0.65cm}
\end{table}

\subsubsection{Consumer Side Experiments}

Consumer side effects can be measured through the standard online randomized experiment set up. The first step in running these experiments is to find an appropriate value of $\alpha$.

\paragraph{Tuning $\alpha$}
$\alpha$ is a parameter (introduced in Equation ~\ref{eq:feedscoredelta}) which controls the balance of the creator and consumer utility. For both the heuristic model and the pCreate models, we first ran some offline simulations to pick a range of $\alpha$ which did not drop the consumer metrics too much. The eventual fine-tuning was done by online A/B tests performed by randomization on the consumer side. The performance numbers reported are based on an $\alpha$ that was picked in this manner.

Once $\alpha$ is decided, we ran a standard A/B test to observe the consumer side effect of both the heuristic model and the parameterized form of feedback sensitivity. For the base heuristic model, $\alpha$ was picked to make the consumer metrics neutral. For the pCreate model, we observed small lifts even in the consumer side metrics with a tuned value of $\alpha$, and also some sensitivity of the metrics with variation in $\alpha$. The results for two values of $\alpha$ from an online experiment are shown in Table \ref{tab:online_consumer}. This indicates the feedback sensitivity is not only positively correlated with consumer metrics of interest, it captures certain aspects of consumer engagement, which our current consumer utilities are not covering. However, it should also be noted that the consumer side wins are small. Also, both the heuristic base model and the pCreate model test boosts the $pFeedback$ term, which is part of the consumer utility as well.


\begin{table}[h]
\centering
\begin{tabular}{l | l | l} 
\hline\hline 
\multirow{2}{*}{Metrics} & \multicolumn{2}{c}{Delta \% Effects  (p-value)} 
\\[0.5ex]
\cline{2-3}
& \multicolumn{1}{c|}{$\alpha_{1}$} & \multicolumn{1}{c}{$\alpha_{2} (> \alpha_{1})$} \\
\hline 
Contributors (Daily)  & $+0.26\%$ ($0.03$) &  Neutral\\
\hline 
Contributors (Weekly)  & $+0.22\%$ ($0.2$) &   Neutral\\
\hline 
Public Contributors (Daily)  & $+0.31\%$ ($0.04$) &  $+0.3\%$ ($0.03$)\\
\hline 
Feed Viral Actors  & $+0.47\%$   ($5e-4$) &  $+0.29\%$ ($0.01$)\\
\hline 
Feed Interactions & Neutral &   $+0.38\%$ ($0.05$)\\
\hline 
\end{tabular}
\caption{Consumer Side Metrics Impact ("Neutral" represents p-value > 0.05)} 
\label{tab:online_consumer}
\vspace{-0.6cm}
\end{table}

\subsubsection{Creator Side Experiments}

To measure creator side impact, we leveraged the "Ego Clusters" experiment set up where the randomization unit is a cluster of users rather than an individual user \cite{Guillaume2019, bonnie2018}. The measurement is only done on the ego member of the cluster. Such a randomization on the clusters enables measurement of the effect from peers/network, but the trade-off is a large reduction on the sample size which leads to:
\begin{enumerate*}
\item Low statistical power and high error-margin in tests.
\item Under-estimation of the treatment effect due to partial overlap in clusters. 
\end{enumerate*}
 
\begin{figure}
  \centering
  \includegraphics[width=0.9\linewidth]{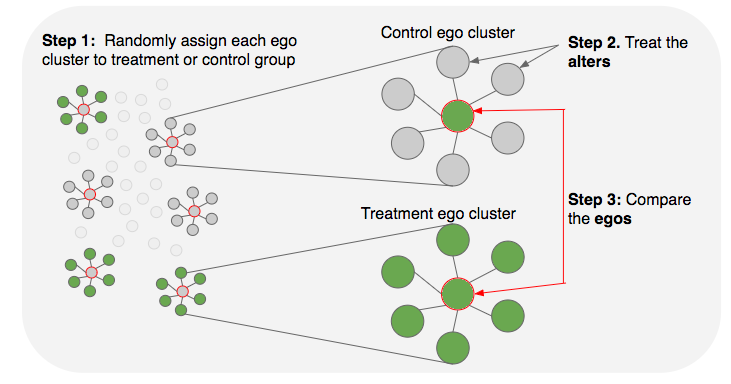}
  \caption{Ego Cluster Set Up}
  \label{fig:ego}
\end{figure}

As Figure \ref{fig:ego} illustrates, in this method, we take a sample of members referred to as the ``Egos'' (following ego network definitions ~\cite{egonet2012}) and randomize them into either the treatment or control group. But instead of enabling the treatment or control experience for the Egos themselves, we treat their Alters (their connections): 
\begin{enumerate*}
\item Feedback sensitivity utilities are incorporated in ranking Alters' feed for the treatment clusters. 
\item Alters' feed ranking uses baseline heuristic model for the control.
\end{enumerate*}

Ego cluster experiments aim to replicate for each Ego, a network which is treated as if the whole population has been treated. This is the rationale behind the described treatment allocation, post randomization of the clusters. Due to the aforementioned limitations of cluster-level randomization, only relatively large impacts can be captured by the "Ego Clusters" experiments with any reasonable statistical significance (i.e., low p-value). Table \ref{tab:online_creator} shows the creator side metrics impact from: 
\begin{enumerate*}
\item Base heuristic model vs control (which had no creator utility term)
\item pCreate model (treatment) vs the base heuristic model (control). 
\end{enumerate*}\

The heuristic model launch showed significant lifts on the creator metrics ($+5.26\%$ retained creators) along with an increase on the feed activities metrics ($+4.99\%$ feed viral actions on mobile). The pCreate model based utilities further improve the creator experience where we observe \textbf{incremental} positive impact on feed activities metrics ($+9.72\%$ feed viral actions on desktop) and an increase in number of contributors who receive a response in time ($+4.9\%$ public contributor with response). Although the significant impacts on overall contributors metric are not yet detected, metrics regarding to heavier type of contribution activities (Weekly Commenter and Total Reshares) have also shown large improvements. We also observed positive incremental lift ({\color{gray}$+3.5\%$}) on Retained Creators (weekly) metric, however the impact is not significant potentially due to the limited statistical power in the network A/B measurement.

The fact that we measured significant positive impact on various critical metrics on the creator side over the baseline model is a strong validation of our feedback sensitivity utilities. We are currently in the process of testing the tree model (XGBoost) based feedback sensitivity utilities online, and expect an even greater amount of metric improvement (based on the offline results previously reported).

\begin{table}[h]
\centering
\begin{tabular}{ p{3.6cm} |  p{2cm} |  p{2cm}} 
\hline\hline 
\multirow{2}{*}{Metrics} & \multicolumn{2}{c}{Delta \% Effects  (p-value)} 
\\[0.5ex]
\cline{2-3}
  & Heuristic  vs  No creator-utility & Pcreate vs Heuristic \\
\hline 
Public Contributors with Response (Daily) &  NA & $+4.9\%$ ($0.03$)\\
\hline 
Retained Creators (Weekly) &  $+5.26\%$ ($0.02$)&\textcolor{gray}{$+3.5\%$ ($0.19$)}\\
\hline 
Feed Viral Actions (mobile) & $+4.99\%$  ($0.01$)& Neutral\\
\hline 
Feed Viral Actions (desktop) & Neutral& $+9.72\%$ ($0.03$) \\
\hline 
Commenter (Weekly)&Neutral& $+2.98\%$ ($0.04$)\\
\hline 
Total Reshares & Neutral& $+12.04\%$ ($0.04$) \\
\hline 
\end{tabular}
\caption{Creator Side Metrics Impact ("Neutral" represents P-values > 0.05)} 
\label{tab:online_creator}
\vspace{-0.6cm}
\end{table}

\section{System Architecture}
\label{sec:architecture}

\begin{figure}
  \centering
  \includegraphics[width=0.98\linewidth]{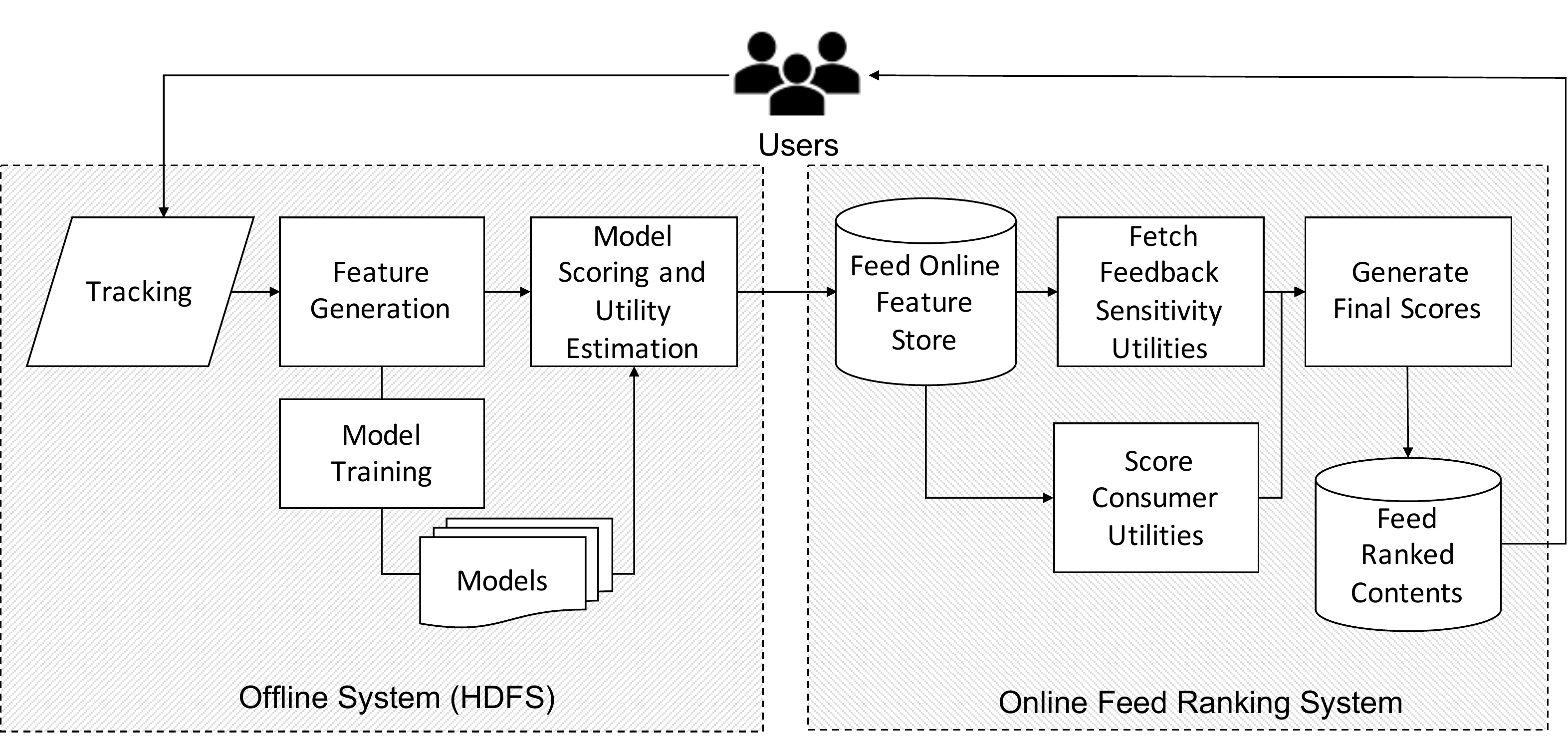}
  \caption{System Architecture}
  \label{fig:systemArchitecture}
\end{figure}

We summarize a design of the system that generates the creator side utility and incorporates it in the feed online recommendation system in Figure \ref{fig:systemArchitecture}. Since feedback sensitivity is relatively static, we build an offline pipeline to generate daily estimates with the following data processing flows:

\begin{enumerate}
\item A flow to fetch and process tracking data and to generate offline features; 
\item Model training pipeline to generate ``pCreate'' models with regularly updated feature data.
\item Scoring flows to calculate the probability of creation based on the features, and calculate the feedback sensitivity utilities ($\delta \hat{p}_{i}$, $\hat{\tau}_{i}$, $\hat{b}_{i}$) as illustrated in Section \ref{sec:estimationOfFeedbackSensitivityUtility}.
\item A job to regularly push updated utility scores computed by the offline flow to an online feature store. 
\end{enumerate}

We process the tracking data, do feature transformation and train the models on Spark. Training is done via Photon library ~\cite{ogilvie2016} for logistic regression and XGBoost library ~\cite{Chen:2016:XST:2939672.2939785} for the gradient boosted tree model. In the online feed recommendation system, various sources of updates (e.g., network's updates, job recommendations, and breaking news) are evaluated via a real-time ranker. The system scores the consumer side utilities using online features and fetches the feedback sensitivity utilities from the online store. It then combines the utilities and generates the final ranking. 

\section{Conclusion}
\label{sec:discussion}

In this paper, we present an approach to better represent creators' interests in a content ecosystem via a consumer facing product, namely newsfeed. There are two key components --- first, building a model to predict creation behavior and estimate feedback sensitivity of each creator, and second, incorporating the personalized creator sensitivities into feed ranking and designing experiments to measure improvements on the creator and consumer side. We showed various offline validation for the precision of our modeling approach, and online validation of the approach which is now fully deployed on the LinkedIn feed and delivering significant wins for creators without compromising the consumer experience.

We are currently testing the tree models in production. One specific extension which seems promising is identifying how creation behavior is affected by different kinds of feedback (i.e., likes, comments, reshares) as well as feedback from different consumers. Another line of future work is more accurate attribution since the causal effect of any specific feedback is unobserved, and possible solutions may involve designing special randomized experiments.

\section*{Acknowledgement}
We would like to thank Preetam Nandy, Rakesh Malladi, Shipeng Yu, Yan Gao and the anonymous KDD reviewers for their careful review and insightful comments. Ajith Muralidharan and Kinjal Basu provided great input during the problem formulation phase. Dylan Wang and the Feed AI team were true partners in the implementation and launch of experiments.

\bibliographystyle{ACM-Reference-Format}
\bibliography{pcreate} 

\section{Reproducibility}

We would like to provide some context on how some of our results can be reproduced or used/adopted otherwise by readers of the paper. In addition, we would like to help readers understand the choices behind not revealing certain information and why that should not deter them from expecting good results in their own endeavors on a similar problem.

\subsection{Detailed implementation plan}
First, we would like to list the various steps in our work and point out which parts have open source tools (or at least algorithms) for the reader to check out.

\begin{enumerate}
\item Analysis tools:
\begin{itemize}
\item Correlation analysis: We processed the tracking data using Apache Spark and further analyzed and visualized the data using Python. 
\end{itemize}
\item Train pCreate model: We used Photon \cite{ogilvie2016} for the logistic regression model, and XGBoost \cite{Chen:2016:XST:2939672.2939785} for the tree model.
\item $\tau$ and $b$ estimation: This can be done at scale by using Equation \ref{eq:tau_intercept} using Spark and developing a few user defined functions with the basic Dataset based implementations.
\item Randomized A/B testing: We conducted experiments and measured impact by using LinkedIn's internal A/B testing platform XLNT \cite{Xu:2015:ICA:2783258.2788602}. The paper describes in depth the experimentation platform and challenges of running A/B tests at large scale social networks. The LinkedIn engineering blog post on XLNT \cite{xlnt} also summarizes the salient features of the platform quite well. Commercial tools such as Optimizely could be an alternative in a general setting. 
\item Ego-cluster: We leveraged LinkedIn's internal tool to measure the creator side impact. This tool was also used in \cite{bonnie2018} and is described in detail in \cite{Guillaume2019}. We would also recommend following the design in \cite{gui2015network} for measuring the peers/network impact. 
\end{enumerate}
\vfill\eject
\subsection{Data sensitivity choices}
We would also like to help the reader better understand our choices behind revealing and concealing certain information. The general principles we have to follow while reporting the performance of any described method on our products include:
\begin{itemize}
\item We cannot compromise a member's privacy. Hence, any data that reveals member information that is not publicly available cannot be reported.
\item We cannot reveal certain business-sensitive information. This generally includes numbers which may have some (potentially remote) connection to revenue or other key metrics, which are not publicly available.
\end{itemize}
Unless otherwise stated, when we choose to reveal any information, it is implied that it passes both the criteria listed above. Given these constraints, here are the decisions we took and some reasoning behind them:
\begin{enumerate}
\item Training data size: We provided details here since the data size is generally critical to the success of the training process.
\item Details of the training data generation: Same rationale as above.
\item Actual feedback sensitivity scores: We decided not to reveal the actual numbers, but provide some relative intuition among large member groups. This was because the actual feedback sensitivity numbers may be too revealing of creator behavior, and also lead to unintentional revelation of some business sensitive metrics.
\item Actual Probability of Creation vs Past Feedback Received: Same rationale as above.
\item Offline Evaluation Metrics: We provided the details on the model evaluation metrics (AUROC and AUPRC), as they are important measures of the model success.
\item Online Metrics: We shared the relative impact (in the form of delta $\%$ effects) but not the absolute delta effects to showcase the success of this approach without revealing sensitive member data statistics.
\end{enumerate}

\end{document}